\documentclass{ws-procs9x6}

\begin{document}

\title{The Bounds on Lorentz and CPT Violating Parameters in the Higgs Sector}

\author{Ismail Turan}

\address{Physics Department, Concordia University, \\
7141 Sherbrooke West, Loyola Campus, \\ 
Montreal, Qc., H4B 1R6  CANADA\\ 
E-mail: ituran@newton.physics.metu.edu.tr}

\maketitle

\abstracts{In this talk, I discuss possible bounds on the Lorentz and CPT violating parameters in the Higgs sector of the so called 
minimal standard model extension. The main motivation to this study is coming from the fact that unlike the parameters in the fermion 
and gauge sector, there are no published bounds on the parameters in the Higgs sector. From the one-loop contributions to the photon 
propagator the bounds on the CPT-even asymmetric coefficients are obtained and the $c_{\mu\nu}$ coefficients in the fermion sector 
determine the bound on the CPT-even symmetric coefficients. The CPT-odd coefficient is bounded from the non-zero vacuum expectation 
value of the $Z$-boson.}

\section{Introduction} 
Lorentz and CPT symmetries are assumed to be exact in nature within the framework of the standard model and this fact is in very good 
agreement to high precision with present-day experimental findings. However, it is widely believed that the standard model is nothing 
but a low energy version of some more complete (fundamental) theory, presumably valid at the Planck scale of $10^{19}$ GeV, such as 
noncommutative field theory\cite{Carroll:2001ws} or string theory.\cite{Kostelecky:1988zi} It is then reasonable to search for some 
induced ``new physics'' 
effects at levels attainable by high precision experiments. The violation of Lorentz and CPT symmetries can be considered one of such 
effects.

There is an explicit example from string theory in which non locality of the string leads to modification of the Lorentz 
properties of the vacuum. Among mechanisms to describe Lorentz and CPT violation, the most elegant way is to consider these symmetries 
exact at the scale of the fundamental theory and spontaneously broken at low energies due to the existence of nonvanishing expectation 
value of some background tensor fields.

The 4-dimensional effective interactions between the background tensor fields $T$ and matter can be written as\cite{Colladay:1996iz}
\begin{eqnarray}
{\it L}^{\prime}\supset\frac{\lambda}{M_{pl}^k}\langle T \rangle\cdot 
\bar{\psi}\Gamma(i\partial)^k\psi+H.c.\,,\;\;k\le 2\label{eq:lag}
\end{eqnarray}
 where all possible Lorentz indices are suppressed. For $k=0,1$, the first two factors of the right hand side of Eq.~(\ref{eq:lag}) 
represent most of the CPT-violating terms in the fermion sector. At this point it is better to explain the difference between the {\it 
observer} Lorentz invariance and the {\it particle} Lorentz invariance, which are essential for understanding the minimal standard 
model-extension (SME) that I will describe briefly in the next section. The former involves transformations under rotations and boosts 
of coordinate system but the latter involves boosts on particle or localized fields but not on the background fields. Therefore, while, 
in the right hand side of Eq.~(\ref{eq:lag}), $\langle T \rangle$ and $\bar{\psi}\Gamma i\partial\psi$ are both changing under {\it 
observer} Lorentz transformation such that their contraction stays invariant, {\it particle} Lorentz transformation leaves $\langle T 
\rangle$ term unaffected which leads to a (particle) Lorentz violating effect when it is contracted with the matter term. The following 
example from conventional electrodynamics\cite{Colladay:1996iz} can be given to give further clarification. Let us consider a charged 
particle entering a region perpendicular to a uniform background magnetic field. Its path is circular. Suppose without changing the 
observer frame, one gives an instantaneous particle boost to the charged particle without effecting its direction. Then it will still 
keep moving on a circular path but with a bigger or smaller radius depending on the direction of the given boost. This boost leaves the 
background magnetic field unaffected (here, the background magnetic field is analogous to the field $T$). Let us now consider another 
observer frame which is obtained from our original frame by making a Lorentz transformation of coordinates. In that frame, the particle 
no longer makes a circular motion but a spiral motion (drift motion) due to the existence of induced electric field in addition to the 
magnetic field. The background field is obviously not a pure magnetic field at all. The important point is that the background field is 
changing to preserve the observer invariance, i.e. $F_{\mu\nu}F^{\mu\nu}$ term is invariant. This means that any Lorentz indices in 
each term of Eq.~(\ref{eq:lag}) must be contracted.

The outline of the talk, which is based on work done with David L. Anderson and Marc Sher\cite{Anderson:2004qi}, is as follows. In Sec. 
2, I will very 
briefly 
describe the minimal standard model extension by especially 
emphasing its fermion, photon and Higgs sector. The purpose of our study is to explore the bounds on the parameters appearing in the 
Higgs sector of the minimal SME. So, in Sec. 3, I consider the bounds on the CPT-even antisymmetric and symmetric coefficients of the 
Higgs sector. A careful analysis of the coordinate and field redefinition issue will be done. The bounds on CPT-odd coefficient in the 
same sector are discussed in Sec. 4.

\section{The minimal Standard Model-Extension}

A framework for studying Lorentz and CPT violation has been constructed by Colladay and Kosteleck\'y\cite{Colladay:1998fq}, known as 
the minimal SME. It is a model based on the standard model but which relaxes the Lorentz and CPT invariance. The additional induced 
terms representing such violation are still invariant under $SU(3)\times SU(2)\times U(1)$ gauge group of the standard model. As 
explained earlier, they preserve the observer Lorentz invariance but not the particle Lorentz invariance. The parameters in the minimal 
SME are assumed to be constant over space-time and this is the reason why we call it ``minimal". An extension of the model by including 
gravity in the context of some non-Minkowski spacetimes has been recently discussed by Kosteleck\'y\cite{Kostelecky:2003fs} and the 
parameters become spacetime dependent.

As an example, for simplicity, the QED sector of the minimal SME which involves the electron and photon sectors is given here.  
\begin{eqnarray} 
{L}_f=\frac{1}{2}i\bar{\psi}\Gamma^{\mu}\stackrel{\leftrightarrow}{D_\mu}\psi-\bar{\psi}M\psi\,, \nonumber 
\end{eqnarray} 
where $\Gamma^{\mu}$ and $M$ denote 
\begin{eqnarray} 
\Gamma^{\mu}&=&\gamma^{\mu}+\Gamma^{\mu}_1\,, \nonumber \\ 
\Gamma^{\mu}_1&\equiv& 
c^{\nu\mu}\gamma_{\nu}+d^{\mu\nu} \gamma_5 \gamma_\mu + e^\mu + i f^\mu \gamma_5 + \frac{1}{2} g^{\lambda\nu\mu} 
\sigma_{\lambda\nu}\nonumber\,, \\ 
M&=&m+M_1\,,\nonumber \\ 
M_1&\equiv& a_\mu \gamma^\mu + b_\mu \gamma_5 \gamma^\mu + \frac{1}{2} H_{\mu\nu} 
\sigma^{\mu\nu}\,.\nonumber 
\end{eqnarray} 
Here all constants $a,b,..,g$ and $H$ represent expectation values of some background tensor 
fields and break the particle Lorentz invariance. The photon sector is given as 
\begin{eqnarray}
 {L}_{p}=- \frac{1}{4} F^{\mu\nu} F_{\mu\nu}- \frac {1}{4} (k_F)_{\kappa\lambda\mu\nu} F^{\kappa\lambda} F^{\mu\nu}+ \frac{1}{2} 
(k_{AF})^{\kappa} \epsilon_{\kappa\lambda\mu\nu} A^\lambda F^{\mu\nu}\,,\label{eq:photlag} 
\end{eqnarray} 
where the Lorentz violation is 
represented by $k_F$ and $k_{AF}$ terms. The parameters have some properties. Let us quote some of them here. All terms in $M_1$ and 
$k_{AF}$ have dimension of mass while all terms in $\Gamma_1^{\mu}$ and $k_F$ are dimensionless. $(k_F)_{\kappa\lambda\mu\nu}$ is 
antisymmetric with respect to first two and last two indices separately and it satisfies the double-trace condition, 
${(k_F)_{\mu\nu}}^{\mu\nu}=0$, to be sure that the photon field is normalized properly. Only $b_\mu,c_\mu$ and 
$(k_F)_{\kappa\lambda\mu\nu}$ will be relevant to our discussion here and there are many experimental and theoretical talks about them 
in this meeting.

The Higgs sector is
\begin{eqnarray} 
{L}_{\rm 
Higgs}&=&(D_{\mu}\Phi)^{\dagger}D^{\mu}\Phi+\mu^2\Phi^{\dagger}\Phi-\frac{\lambda}{3!}(\Phi^{\dagger}\Phi)^2
+{L}_{\rm Higgs}^{\prime \rm CPT-even}+{L}_{\rm Higgs}^{\prime \rm CPT-odd}\,,\nonumber \\ 
{ L}_{\rm Higgs}^{\prime {\rm 
CPT-even}}&=&\left[\frac{1}{2}(k_{\phi\phi}^{S}+ik_{\phi\phi}^{
    A})_{\mu\nu}(D^{\mu}\Phi)^{\dagger}D^{\nu}\Phi +{\rm H.c.}\right]
-\frac{1}{2} k_{\phi
    B}^{\mu\nu}\Phi^{\dagger}\Phi B_{\mu\nu}\nonumber\\ 
&&-\frac{1}{2}k_{\phi W}^{\mu\nu}\Phi^{\dagger}W_{\mu\nu}\Phi\,,\nonumber \\ 
{L}_{\rm Higgs}^{\prime \rm CPT-odd}&=& i k_{\phi}^{\mu}\Phi^{\dagger}D_{\mu}\Phi + {\rm H.c.}\,,\nonumber 
\end{eqnarray} 
where $k_{\phi\phi}$ has real symmetric and imaginary antisymmetric parts, which are separated as above and $k_{\phi B}$ 
and $k_{\phi W}$ have 
only real symmetric parts. All are CPT preserving (CPT-even) but Lorentz violating and dimensionless. The only CPT-odd and mass 
dimension coefficient is $k_\phi$. 

\section{The CPT-even coefficients}
\subsection{The CPT-even antisymmetric coefficients }
Direct detection of these coefficients
would necessitate producing large numbers of Higgs bosons, and the 
resulting bounds would be quite weak.  However, there are extremely 
stringent bounds on Lorentz violation at low energies, and 
thus searching for the effects of these new interactions through loop 
effects will provide the strongest bounds. The most promising of these effects 
will be on the photon propagator.

In this section, we will consider the bounds on the CPT-even antisymmetric coefficients, $k_{\phi\phi}^A, k_{\phi B}$ and $k_{\phi 
W}$.We first look at the most general CPT-even photon propagator, and then relate the $k_{\phi\phi}^A$ coefficients to the 
Lorentz-violating terms in the photon propagator.  Then, the experimental constraints on such terms lead directly to stringent bounds on 
the $k_{\phi\phi}^A$ coefficients.  We then consider the $k_{\phi B}$ and $k_{\phi W}$ coefficients.

The equation of motion from the Lagrangian Eq.~(\ref{eq:photlag}) is\footnote{We set $k_{AF}$-term to zero, since it is very tightly 
constrained from astrophysical observations.\cite{Colladay:1998fq}} 
\begin{equation} 
M^{\alpha\delta}A_{\delta}=0\,,\end{equation} 
where \begin{equation} M^{\alpha\delta}(p) \equiv g^{\alpha\delta}p^{2}-p^{\alpha}p^{\delta}-2 
(k_{F})^{\alpha\beta\gamma\delta}p_{\beta}p_{\gamma}\,.\label{photprop} \end{equation} 

The propagator is clearly gauge invariant (recall 
that $k_{F}$ is antisymmetric under exchange of the first or last two indices). Note that while the $g^{\mu\nu}p^{2}-p^{\mu}p^{\nu}$ 
structure is mandated by gauge invariance, the $k_{F}$ term is separately gauge invariant and may differ order by order in perturbation 
theory. For simplicity, we look at the divergent parts of the one loop diagrams only. Consideration of higher orders and finite parts 
will give similar, although not necessarily identical, results. We can consider each of the possible terms independently by assuming 
that there is no high-precision cancellations. Let us start with $k_{\phi\phi}^A$.

The $k_{\phi\phi}^A$-term leads to photon-Goldstone boson-$W$ boson and photon-Goldstone boson-Goldstone boson type interactions which 
are absent in the conventional Standard Model. As we do in the Standard Model, it is possible to fix the gauge to simplify the 
calculations. The Standard Model gauge fixing removes the mixing between $W^{\mp}$ boson and the charged Goldstone boson $\phi^{\pm}$. A 
similar situation happens in the minimal SME if one modifies the gauge fixing functions by adding a 
$i(k_{\phi\phi}^{A})_{\mu\nu}\partial^{\mu}A_i^{\nu}$ term to the $SU(2)$ functions and a similar 
$i(k_{\phi\phi}^{A})_{\mu\nu}\partial^{\mu}B^{\nu}$ to the $U(1)$ function.\cite{Anderson:2004qi} However, such 
generalization also 
leads to an unwanted mixing between the gauge boson $Z_{\mu}$ and the derivative of the Higgs field, $\partial_{\nu}\phi_1$, which is 
contracted with $(k_{\phi\phi}^{A})^{\mu\nu}$, as well as substantially complicating the photon propagator. An easier way is to use a 
mixed propagator of the form 

\begin{figure}[ht] 
\vspace*{-0.2cm} 
\centerline{\epsfxsize=2.3in\epsfbox{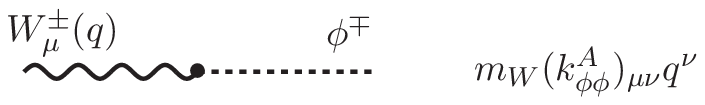}} 
\label{mixing} 
\end{figure}

\begin{figure}[ht]
\vspace*{1.5cm}
\centerline{\epsfxsize=2.8in\epsfbox{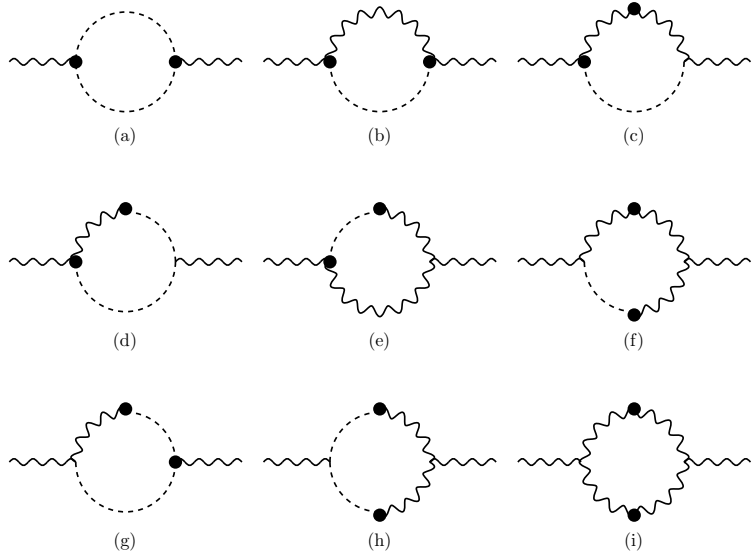}}   
\caption{One-loop contributions to the photon vacuum polarization
involving Lorentz-violating
interactions to second order. These diagrams are for
$k_{\phi\phi}^A$ case but similar diagrams
exist for the other antisymmetric coefficients. Here the wavy (dashed)
line circulating in the loop represents $W$ boson (charged Goldstone
boson).
Each blob in vertices, $W$-propagator or $W\!\!-\!\!\phi$ mixed
propagator represents a single Lorentz-violating coefficient insertion. 
The rest of the diagrams can be obtained by
permutations of these 9 diagrams.}\label{oneloop}
\end{figure}

Another feature of the $k_{\phi\phi}^A$-term is the modification of the $W$-boson propagator. 
Up to the second order in $k_{\phi\phi}^A$, the propagator in the 't Hooft-Feynman gauge takes the form
\begin{eqnarray}    
i\Delta_{\nu\lambda}(\xi=1)&=&i\Delta^{(0)}_{\nu\lambda}+m_W^2\frac{(k^A_{\phi\phi})_{\nu\lambda}}{(q^2-m_W^2)^2}+im_W^4
\frac{(k^A_{\phi\phi})_{\nu\alpha}{(k^A_{\phi\phi})^{\alpha}}_{\lambda}}{(q^2-m_W^2)^3}\,.
\label{propW}
\end{eqnarray}
The one-loop contributions to the photon vacuum polarization are given in Fig.~\ref{oneloop}. 
Here we only include diagrams with second order $k_{\phi\phi}^A$, 
since one can show that all diagrams with one $k_{\phi\phi}^A$ inclusion vanish.

There are two possible structures in second order, which are either $(k^A_{\phi\phi})_{\mu\lambda}{(k^A_{\phi\phi})^{\lambda}}_{\nu}$ or 
$(k^A_{\phi\phi})_{\mu\lambda} (k^A_{\phi\phi})_{\lambda^{\prime}\nu}p^{\lambda}p^{\lambda^{\prime}}$. Here $p$ is the four momentum of 
the external photons. Again the first possibility is not gauge invariant and should vanish, thus contributions from the third term in 
Eq. (\ref{propW}) should vanish.  We have verified this explicitly.  The latter is gauge invariant and gives a non-zero contribution (if 
we contract with any of two external momenta of photons, $p^{\mu}$ or $p^{\nu}$, it vanishes due to the antisymmetry property of 
$k^{A}_{\phi\phi}$). Calculating the one-loop diagrams, and  comparing
with Eq. (\ref{photprop}), we find that the components of $k_F$ can simply be expressed in terms of 
$k^A_{\phi\phi}$ as 
$(k_F)_{\mu\lambda\lambda^{\prime}\nu}=\frac{1}{3}(k^A_{\phi\phi})_{\mu\lambda}
(k^A_{\phi\phi})_{\lambda^{\prime}\nu}$.   
We now turn to the experimental bounds on the $k_{F}$. Many speakers in this meeting have talked about the 
$\tilde{\kappa}_{e+}$ and by $\tilde{\kappa}_{o-}$ which are $3\times 3$ matrices defined from the components of 
$k_F$ and represent 10 
of 19 elements of $k_F$.\cite{kostelecky:2002hh} The strongest bound is coming from birefringence constraints\cite{kostelecky:2002hh} 
and is given by 
$3\times 10^{-32}$. I should note  that for any single or possible combination 
of non-zero elements of $(k_{\phi\phi}^A)_{\mu \nu}$ it is impossible 
for both
 $\tilde{\kappa}_{e+}$ and $\tilde{\kappa}_{o-}$ to be null matrices, 
 and thus the birefringence constraints apply.\cite{Anderson:2004qi} Therefore the upper bound of the 
$(k_{\phi\phi}^A)_{\mu \nu}$ coefficients can be obtained as $3\times 10^{-16}$.

The discussion of $k_{\phi B}^{\mu\nu}$ and $k_{\phi W}^{\mu\nu}$ is very parallel to the $k_{\phi\phi}^A$ case. $k_{\phi B}^{\mu\nu}$ 
term does not induce a $W$-Goldstone mixing but leads to photon-Higgs scalar mixing instead. The $k_{\phi W}^{\mu\nu}$ term has very 
similar 
features to the $k_{\phi\phi}^A$ case except for the photon-Higgs-boson mixing. The
$(k_F)_{\mu\lambda\lambda^{\prime}\nu}= 
\frac{5}{12e^2}\cos^2{\theta_W}(k_{\phi B})_{\mu\lambda}(k_{\phi 
B})_{\lambda^{\prime}\nu}$ and   $(k_F)_{\mu\lambda\lambda^{\prime}\nu} = -\frac{5}{12e^2}\sin^2{\theta_W}(k_{\phi 
W})_{\mu\lambda}(k_{\phi W})_{\lambda^{\prime}\nu}$ equalities hold, which sets the bound as $0.9\times 10^{-16}$ and $1.7\times 
10^{-16}$, respectively. It is seen that the current bound on all three Lorentz violating coefficients is of the order of $10^{-16}$ and 
can easily be updated as the bound on $k_{F}$ is updated.

\subsection{Coordinate and field redefinitions and the symmetric coefficients}
We consider bounds on the $k_{\phi\phi}^S$ coefficients.  In this case, the strongest
bounds come from relating, through field redefinitions, these coefficients to other Lorentz violating
coefficients in the fermion sector, and then using previously determined bounds on those coefficients.
Therefore, Coordinate and field redefinitions need to be discussed carefully.

Once any model is extended by relaxing some its symmetry properties, not all of the new parameters representing an apparent violation 
of 
these symmetries may be physical. That is, the model has some redundant parameters. Therefore, the extended model  should be carefully 
analyzed 
to check for redundant parameters. This analysis may yield several Lagrangians which are equivalent to each other by some coordinate and 
field redefinitions and rescalings\cite{Colladay:1998fq,Colladay:2002eh,Colladay:2003tj,Muller:2004mb}. The same situation applies to 
the minimal SME case. A simple observation from the fermion sector is that 
$\overline{\psi}\gamma^{\mu}D_{\mu}\psi-a_{\mu}\overline{\psi}\gamma^{\mu}\psi\rightarrow \overline{\psi}\gamma^{\mu}D_{\mu}\psi$ under 
$\psi\rightarrow exp(-ia^{\mu}x_{\mu})\psi$. Thus, $a_{\mu}$ is redundant unless gravity is included. A similar conclusion can be drawn 
for some components of $k_{\phi\phi}^S$ in the Higgs sector under certain circumstances. 
Consider a case\cite{kostelecky:2002hh,Muller:2004mb} with only two Lorentz-violating parameters $k_{\phi\phi}$ and $k_F$ in the 
scalar and photon sectors, respectively. The Lagrangian is ${L}=\left[g_{\mu\nu}+(k_{\phi\phi})_{\mu\nu}\right](D^{\mu}\Phi)^{ 
\dagger}D^{\nu}\Phi-m^2\Phi^{\dagger}\Phi-\frac{1}{4}F_{\mu\nu}F^{\mu\nu}-\frac{1}{4}(k_F)_{\mu\lambda\lambda^{\prime}\nu} 
F^{\mu\lambda}F^{\lambda^{ \prime}\nu}$, where $D_{\mu}=\partial_{\mu}+iqA_{\mu}$ and $k_{\phi\phi}$ is real and symmetric. Let 
us assume that only one component of $k_{\phi\phi}$, $(k_{\phi\phi})_{00}\!\equiv\! k^2\!-\!1$, is nonzero\cite{kostelecky:2002hh,Muller:2004mb} and that $k_{F}$ is taken as zero. The transformations $t\to k t$, ${\bf x}\to {\bf x}$ and the 
field redefinitions $A_0\to A_0$, ${\bf A}\to k{\bf A}$ with rescaling of the electric charge $q\to q/k$ move the Lorentz violation into 
the photon sector (${L}_{\rm photon}=(D_{\mu}\Phi)^{\dagger}D^{\mu}\Phi-m^2\Phi^{\dagger}\Phi+\frac{1}{2}(E^2-k^2 B^2)$, where 
$E(B)$ is the electric(magnetic) field). One further example is the following: Consider only 
$(k_{\phi\phi})_{11}=(k_{\phi\phi})_{22}=(k_{\phi\phi})_{33}=\!k^2\!-\!1$ nonzero and then it is still possible to get an equivalent 
Lagrangian as ${L}_{\rm photon}=(D_{\mu}\Phi)^{\dagger}D^{\mu}\Phi-m^2\Phi^{\dagger}\Phi+\frac{1}{2}(E^2-B^2/k^2)$ under the
 transformations $t\to t$, ${\bf x}\to k {\bf x}$ and the redefinitions 
$A_0\to k A_0$, ${\bf A}\to {\bf A}$ with the same 
 charge rescaling $q\to q/k$. However, for the other components of  
$k_{\phi\phi}$, there are no such obvious transformations.

Another observation is from the electron sector of the extended QED\cite{Colladay:2002eh}. The free electron Lagrangian with explicit 
Lorentz 
violation ${L}_{f}^0(\psi(x))$ transforms into ${L}_{f}^0(\chi(x))+ 
ic_{\mu\nu}\bar{\chi}\gamma^{\mu}\partial^{\nu}\chi={L}_{f}^0(\chi(x^{\prime}))$ under the transformation $\psi(x)=(1+c_{\mu 
\nu}x^{\mu}\partial^{\nu})\chi(x)\,\,$ (i.e., $x^{\mu}\to x^{\prime \mu}=x^{\mu}+c_{\nu}^{\mu}x^{\nu}$). Note that $c_{\mu\nu}$ is 
redundant unless fermion-photon interactions are included. Similarly the field redefinition of the Higgs doublet 
$\Phi(x)=\left(1+\frac{1}{2}(k_{\phi\phi}^S)_{\mu\nu}x^{\mu}\partial^{\nu} \right)\varphi(x)$ eliminates the explicit Lorentz violation 
in the Higgs sector but the $(k_{\phi\phi}^S)$-term reappears as a $c$-term in the photon sector. Thus, the redundancy of the parameters 
in 
the minimal SME is a matter of convention. Assuming a conventional fermion sector(and the photon sector in the case of including 
fermion-photon interactions) makes the $(k_{\phi\phi}^S)_{\mu\nu}$ physical. Otherwise, there is mixing among $k_{\phi\phi}^S, 
c_{\mu\nu}$, and nine unbounded $k_F$ coefficients. In this study, we {\it only} concentrate on the Lorentz and CPT 
violation in the scalar sector of the SME, hence we assume that the theory has a conventional fermion sector, which means that bounds on 
$c_{\mu\nu}$ will lead to effective bounds on $k_{\phi\phi}^{S}$. The best current bounds on the components of $c_{\mu\nu}$ are 
summarized in Table \ref{Table1} as direct  bounds on the components of $(k_{\phi\phi}^{S})_{\mu\nu}$. In general, we prefer using 
the 
measured cleaner bounds, if available, to some projected tighter bounds estimated from some planned experiments.

\section{The CPT-odd coefficient}
One interesting effect of the CPT-odd $k_{\phi}$-term is the modification of the conventional 
electroweak $\rm SU(2)\times U(1)$ symmetry breaking. Minimization of the static potential yields a nonzero expectation value for 
$Z_{\mu}$ boson field of the form $\langle 
Z_{\mu}\rangle_{0}={\sin{2\theta_W}\over q} {\rm Re} (k_{\phi})_{\mu}$.\footnote{Here we have assumed all the other Lorentz-violating 
coefficients zero.} The nonzero expectation value for the $Z$ will, 
when plugged into the conventional fermion-fermion-$Z$ interaction, yield a 
$b_{\mu}\overline{\psi}\gamma^{\mu}\gamma_{5}\psi$ term.\footnote{Alternatively, one can look at the one-loop effects on the photon 
propagator, however this will yield much weaker bounds.} Then the relation $b_{\mu}=\frac{1}{4}Re(k_{\phi})_{\mu}$ holds.  The best 
bound on $b_\mu$ for its $X$ and $Y$ components comes from the neutron with the use of a two-species noble-gas maser\cite{Bear:2000cd} 
and it is of 
 the order of $b_{X,Y}^n\leq 10^{-32}$ GeV. Note that in order to get this bound there are some assumptions about the nuclear 
 configurations, which make the bound uncertain accuracy to within one or two orders of magnitude. Details of the 
experiment and some new improvements can be found in the proceedings of both this and the previous meetings. The best bound on the 
$Z$ component of $b_\mu$ comes from testing of cosmic 
spatial isotropy for 
 polarized electrons\cite{Hou:2003} and it is of the order of $b_Z^e\leq 7.1\times10^{-28}$ GeV in the Sun-centered frame. The bound on 
the time component
 of $b_{\mu}$ is around $b_T^n\leq 10^{-27}$ GeV\cite{Cane:2003wp}. The complete list of all bounds on the Lorentz and CPT violating 
parameters of the Higgs sector is given in Table \ref{Table1}. 

  
\begin{table}[ph]
\caption{Estimated upper bounds for the Lorentz and CPT violating coefficients in the Higgs sector of the SME.}
\label{Table1}

\begin{center}

{\footnotesize
    \begin{tabular}{ccccc}
    \hline
Parameters  &&$\;$ Sources & &$\;\;$Comments\\ \hline
	 &$\;\;\;\;\;\;\;\;$$\tilde{\kappa}_{e^{+}},\tilde{\kappa}_{o^{-}}$    & $c_{\mu\nu}$ & $b_{\mu}$ (GeV)&  \\ \hline 
    $(k_{\phi\phi}^A)_{\mu\nu}$    & $\;\;\;\;\;\;\;\;3\times 10^{-16}$      &    -   &   -  &             -   \\
    $(k_{\phi B})_{\mu\nu}$    & $\;\;\;\;\;\;0.9\times  10^{-16}$    &    -  &   - &               -   \\
    $(k_{\phi W})_{\mu\nu}$   & $\;\;\;\;\;\;\;\;1.7\times 10^{-16}$     &    - &  -  &            -   \\
    $(k_{\phi\phi}^S)_{II}$   & $\;\;\;\;\;\;$ -    &   $10^{-27}$&  - &    \   a        \\
    $ (k_{\phi\phi}^S)_{TT}$  & $\;\;\;\;\;\;$ - &$4\times 10^{-13}$& - &   \   b                            \\
    $(k_{\phi\phi}^S)_{TI}$    & $\;\;\;\;\;\;$ -  &    $10^{-25}$& - &  \      c                                    \\
    $(k_{\phi\phi}^S)_{XZ},(k_{\phi\phi}^S)_{YZ}$   & $\;\;\;\;\;\;$ -  &    $10^{-25}$&  - &  d \\
    $(k_{\phi\phi}^S)_{XY}$  & $\;\;\;\;\;\;$ - & $10^{-27}$ & - &          d           \\
    $(k_{\phi})_X, (k_{\phi})_Y$   & $\;\;\;\;\;\;$ -   &    -  &   $10^{-31}$ &  e          \\
    $(k_{\phi})_Z,(k_{\phi})_T$   & $\;\;\;\;\;\;$ -  &    -  &   $2.8\times 10^{-27}$  &    f                  \\
    \hline 
        \end{tabular}}
\begin{minipage}{11cm}
\begin{tabnote}
\item a) Obtained from $c_{\mu\nu}^{\rm neutron}$ with the
assumption that Lorentz violation is not isotropic.\cite{Lamor,Bluhm:2003un,Kostelecky:1999mr} If it is isotropic,the bound on 
$(k_{\phi\phi}^S)_{TT}$ applies.\cite{Gabrielse:1998ee}
\item b) Obtained from the comparison of the anti-proton's frequency with the hydrogen ion's frequency.\cite{Gabrielse:1998ee}
\item c) Estimated value based on the sensitivity calculations of some planned 
space-experiments.\cite{Bluhm:2003un,kostelecky:2003cr,kostelecky:2003xn,Datta:2003dg}
\item d) Obtained from the neutron.\cite{Prest,Lamor,Bluhm:2003un,Kostelecky:1999mr}
\item e) From $b_{\mu}^{\rm neutron}$ with the use of a two-species noble-gas maser. From  $b_{\mu}^{\rm electron}$, a weaker but 
cleaner bound of $1.2\times 10^{-25}$ can be obtained.
\item f) This bound is from the spatial isotropy test of polarized electrons.
\end{tabnote}
\end{minipage}
        \end{center}
\end{table}

\section*{Acknowledgments}
I thank my collaborators Marc Sher and David L. Anderson. I am grateful to V. Alan Kosteleck\'y for many discussions and
encouragement. I am also thankful to Mariana Frank who made this presentation possible.

\end{document}